# Steganography Algorithm to Hide Secret Message inside an Image

Rosziati Ibrahim and Teoh Suk Kuan

*Faculty of Computer Science and Information Technology, University Tun Hussein Onn Malaysia (UTHM), Batu Pahat 86400, Johor, Malaysia*

**Abstract:** In this paper, the authors propose a new algorithm to hide data inside image using steganography technique. The proposed algorithm uses binary codes and pixels inside an image. The zipped file is used before it is converted to binary codes to maximize the storage of data inside the image. By applying the proposed algorithm, a system called Steganography Imaging System (SIS) is developed. The system is then tested to see the viability of the proposed algorithm. Various sizes of data are stored inside the images and the PSNR (Peak signal-to-noise ratio) is also captured for each of the images tested. Based on the PSNR value of each images, the stego image has a higher PSNR value. Hence this new steganography algorithm is very efficient to hide the data inside the image.

## 1. Introduction

This paper proposes a new algorithm to hide the data inside images using steganography technique. An algorithm is designed to hide all the data inputted within the image to protect the privacy of the data. Then, the system is developed based on the new steganography algorithm. This proposed system provides an image platform for user to input image and a text box to insert texts. Once the proposed algorithm is adapted, user can send the stego image to other computer user so that the receiver is able to retrieve and read the data which is hidden in the stego image by using the same proposed system. Thus, the data can be protected without revealing the contents to other people.

Steganography Imaging System (SIS) is a system that is capable of hiding the data inside the image. The system is using 2 layers of security in order to maintain data privacy. Data security is the practice of keeping data protected from corruption and unauthorized access. The focus behind data security is to ensure privacy while protecting personal or corporate data. Privacy, on the other hand, is the ability of an individual or group to seclude them or information about themselves and thereby reveal them selectively. Data privacy or information privacy is the relationship between collection and dissemination of data, technology, the public expectation of privacy, and the legal issues.

Data privacy issues can arise from a wide range of sources such as healthcare records, criminal justice investigations and proceedings, financial institutions and transactions, biological traits, residence and geographic records and ethnicity. Data security or data privacy has become increasingly important as more and more systems are connected to the Internet. There are information privacy laws that cover the protection of data or information on private individuals from intentional or unintentional disclosure or misuse. Thus, hiding the data in a kind of form such as within an

---

**Corresponding author:** Rosziati Ibrahim, Ph.D., research fields: image processing, software specification. E-mail: rosziati@uthm.edu.my.
Suk Kuan Teoh, student, research fields: image processing, software engineering.

image is vital in order to make sure that security or privacy of the important data is protected.

The rest of the paper is organized as follows. Section 2 reviews the related work and section 3 presents the proposed algorithm. The implementation of the system is discussed in section 4 together with the discussion of various results obtained from testing the system based on the proposed algorithm with various sizes of data. The image is also tested using the PSNR value. Finally, we conclude the paper in section 5.

## 2. Related Work

Hiding data is the process of embedding information into digital content without causing perceptual degradation [1]. In data hiding, three famous techniques can be used. They are watermarking, steganography and cryptography. Steganography is defined as covering writing in Greek. It includes any process that deals with data or information within other data. According to Lou et al. [2], steganography is hiding the existence of a message by hiding information into various carriers. The major intent is to prevent the detection of hidden information.

Research in steganography technique has been done back in ancient Greek where during that time the ancient Greek practice of tattooing a secret message on the shaved head of a messenger, and letting his hair grow back before sending him through enemy territory where the latency of this communications system was measured in months [3]. The most famous method of traditional steganography technique around 440 B.C. is marking the document with invisible secret ink, like the juice of a lemon to hide information. Another method is to mark selected characters within a document by pinholes and to generate a pattern or signature [3]. However, the majority of the development and use of computerized steganography only occurred in year 2000 [4]. The main advantage of steganography algorithm is because of its simple security mechanism. Because the steganographic message is integrated invisibly and covered inside other harmless sources, it is very difficult to detect the message without knowing the existence and the appropriate encoding scheme [5]. There are several steganography techniques used for hiding data such as batch steganography, permutation stehanography, least significant bits (LSB), bit-plane complexity segmentation (BPCS) and chaos based spread spectrum image steganography (CSSIS).

Research in hiding data inside image using steganography technique has been done by many researchers, for example in [6-10]. Warkentin et al. [6] proposed an approach to hide data inside the audiovisual files. In their steganography algorithm, to hide data, the secret content has to be hidden in a cover message. El-Emam [7], on the other hand, proposed a steganography algorithm to hide a large amount of data with high security. His steganography algorithm is based on hiding a large amount of data (image, audio, text) file inside a colour bitmap (bmp) image. In his research, the image will be filtered and segmented where bits replacement is used on the appropriate pixels. These pixels are selected randomly rather than sequentially. Chen et al. [8] modified a method used in [9] using the side match method. They concentrated on hiding the data in the edge portions of the image. Wu et al. [10], on the other hand, used pixel-value differencing by partitioning the original image into non-overlapping blocks of two consecutive pixels.

This research uses a similar concept introduced by El-Emam [7]. A bitmap (bmp) image will be used to hide the data. Data will be embedded inside the image using the pixels. Then the pixels of stego image can then be accessed back in order to retrieve back the hidden data inside the image. Two stages are involved. The first stage is to come up with a new steganography algorithm in order to hide the data inside the image and the second stage is to come up with a decryption algorithm using data retrieving method in order to retrieve the hidden data that is hided within the stego image.

## 3. Proposed Algorithm

Our proposed algorithm is using two layers of security to maintain the privacy, confidentiality and accuracy of the data. Fig. 1 shows the framework for the overall process of the system. The system is able to hide the data inside the image as well as to retrieve the data from the image.

From Fig. 1, for hiding the data, a username and password are required prior to use the system. Once the user has been login into the system, the user can use the information (data) together with the secret key to hide the data inside the chosen image. Using a novel steganography algorithm, these data will be embedded and hided inside the image with almost zero distortion of the original image.

For retrieving the data, a secret key is required to retrieving back the data that have been embedded inside the image. Without the secret key, the data cannot be retrieved from the image. This is to ensure the integrity and confidentiality of the data.

For the steganography algorithm, Fig. 2 shows the algorithm for embedding the secret message inside the image. During the process of embedding the message inside the image, a secret key is needed for the purpose of retrieving the message back from the image.

From Fig. 2, the secret message that is extracted from the system is transferred into text file first. Then the text file is compressed into the zip file. The zip text file then is used for converting it into the binary codes.

The purpose of zipping the text file is because the zipped text file is more secured if compared with the file that is without the zipped. The contents in the zipped file will significantly hard to be detected and read. Furthermore, this series of binary codes of the zipped text file and the key is a long random codes in which they only consist of one and zero figures. A data hiding method is applied by using this series of binary codes. By applying the data hiding method, the last two binary codes from the series are encoded into a pixel in image, then, next two binary codes are encoded to the next pixel in image, the process is repeated until all the binary codes are encoded. The secret key in this proposed steganography algorithm is playing an essential role where the key is acts as a locker that used to lock or unlock the secret message. For the data hiding method, each last two bit is encoded into each pixel in image. This will ensure the original image will not be tempered with too many changes.

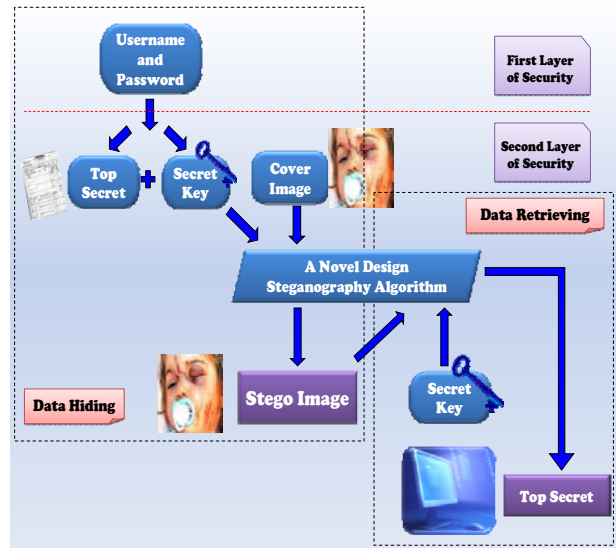

**Fig. 1** The framework for the system.

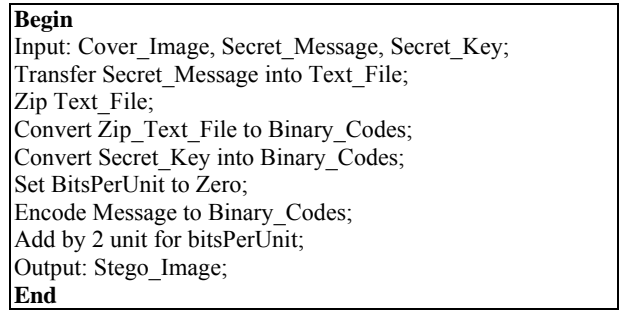

**Fig. 2** Algorithm for embedding data inside image.

Once the message is hidden inside the image, this message can be extracted back from the stego image. Fig. 3 shows the algorithm for extracting the secret message from the stego image. In order to retrieve a correct message from the image, a secret key is needed for the purpose of verification.

From Fig. 3, for the data extracting method, a secret key is needed to detect whether the key is match with the key that decodes from the series of binary code. Once the key is matched, the process continues by forming the binary code to a zipped text file, unzip the text file and transfer the secret message from the text file to retrieve the original secret message.

```
Begin
Input: Stego_Image, Secret_Key;
Compare Secret_Key;
Calculate BitsPerUnit;
Decode All_Binary_Codes;
Shift by 2 unit for bitsPerUnit;
Convert Binary_Codes to Text_File;
Unzip Text_File;
Output Secret_Message;
End
```

**Fig. 3  Algorithm for extracting data from stego image.**

The main focuses of this proposed steganography algorithm are the use of transferring secret message to a text file, zipping file, a key, converting both zipped file and key into a series of binary codes, and the use of encoding each last two binary codes into pixels in image. The image quality is still robust where the distortion and colour changes of images are reduced to the minimum or zero-distortion. Secret message, on the other hand, is difficult to be stolen by steganalysis.

The proposed steganography algorithm consists of two image embedding techniques which are data hiding method and data retrieving method. Data hiding method is used to hide the secret message and the key in cover image while data retrieving method is used to retrieve the key and the hidden secret message from stego image. Hence, data or in particular a secret message, is protected in image without revealing to unauthorized party.

Both from Figs. 2-3 show that 2 layers of security are maintain within the system. However, the secret key is used for verification process in order to retrieve the correct message back from the image. This secret key is also embedded together with the data inside the image. Therefore, when a user is transmitting the image via the internet, that image contains the data and the secret key as well. However, the data can only be retrieved from the image using the system.

## 4. Result and Discussion

Based on the proposed algorithm, we develop a simple system, which implements the algorithm. We name the system as Steganogrphy Imaging System (SIS). Based on the framework for the system as seen in Fig. 1, SIS imposed on 2 layers of security. The first layer is for the login purpose and the second layer is for the hiding and retrieving purposes. The system is introduced in [11]. Fig. 4 shows the main interface for the system.

From Fig. 4, SIS has two main boxes, one box for the image and another box for the data that the user needs to hide inside the image. The image box is used for getting the image from any location and the text box is used for hiding and retrieving the message to and from image respectively. In order to hide the data inside the image, a secret key is required for the purpose of security reason. Fig. 5 shows the interface for the secret key which needs to be in 6 characters.

From Fig. 5, the secret key is required to enter twice for the verification purposes. For simplicity, 6 characters are used for the secret key. This secret key is also embedded inside the image together with the data. Therefore, to reduce the size of storing the secret key inside the image, only 6 characters are used for the secret key. Once the data has been key in and the secret key has been entered, the new stego image can be saved to a different image file. This new stego image can then be used by user to send it via internet or email to other parties without revealing the secret data inside the image. If the other parties want to reveal the secret data hidden inside the image, the new stego image file can then be upload again using the system to retrieve the data that have been locked inside the image using the secret key.

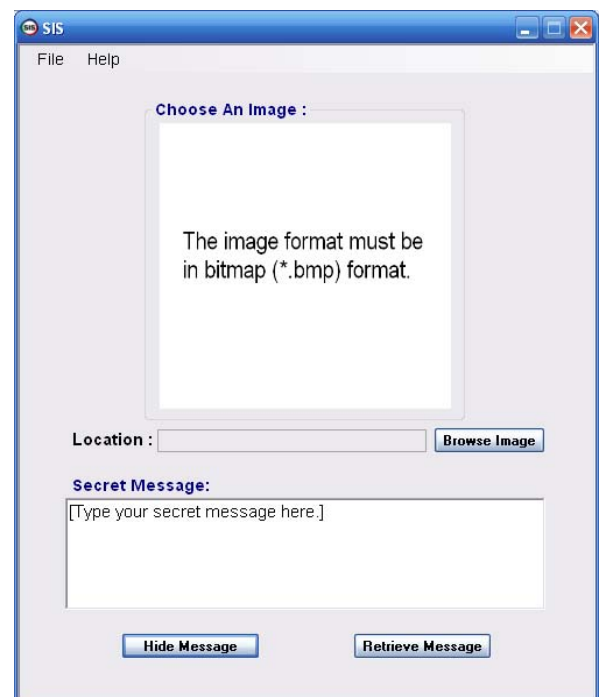

**Fig. 4  The main interface for SIS.**

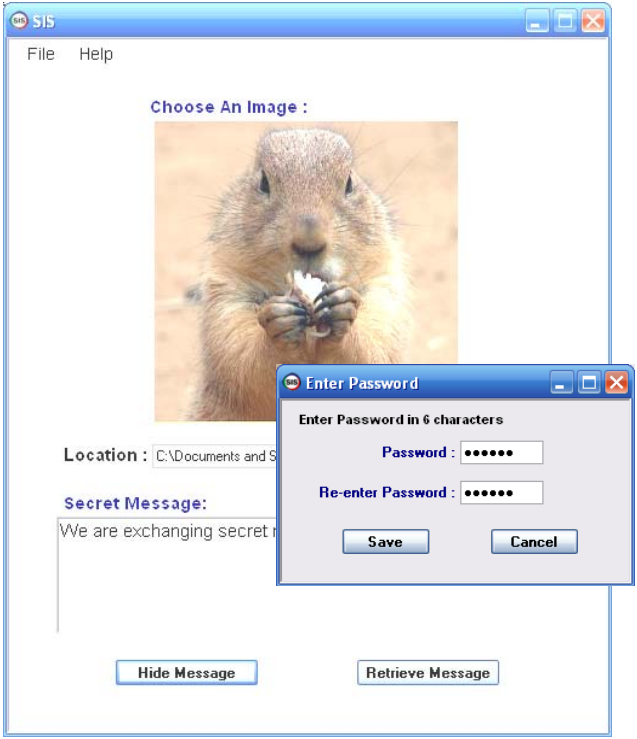

**Fig. 5   The secret key is required for SIS.**

The system is tested using the images as showed in Figs. 6-7. Fig. 6 (a) shows the original image before the message is stored inside the image and Fig. 6 (b) shows the stego image after the message is stored inside the image. We found that the stego image does not have a noticeable distortion on it (as seen by the naked eyes).

Fig. 7 shows another example of image with data hidden inside the image.

From Fig. 7, it shows that the comparison of distortion by naked eyes between cover image and stego image is almost zero. The surfaces of between both images show no difference by using naked eyes even though the size of stego image has a slightly higher than the cover image.

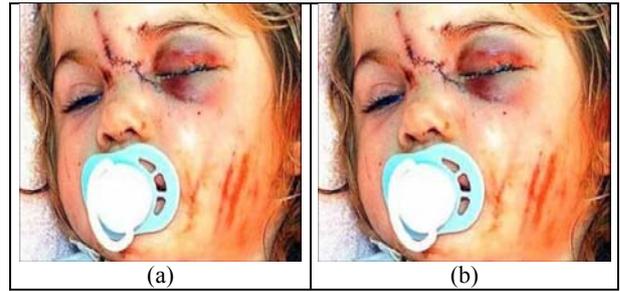

**Fig. 6   (a) Original image    (b) Stego image.**

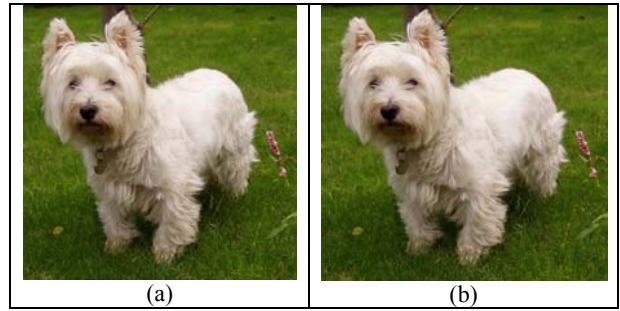

**Fig. 7   (a) Original image    (b) Stego image.**

We then tested the algorithm using the PSNR (Peak signal-to-noise ratio). PSNR is a standard measurement used in steganograpy technique in order to test the quality of the stego images. The higher the value of PSNR, the more quality the stego image will have.

If the cover image is $C$ of size $M \times M$ and the stego image is $S$ of size $N \times N$, then each cover image $C$ and stego image $S$ will have pixel value $(x, y)$ from $0$ to $M-1$ and $0$ to $N-1$ respectively. The PSNR is then calculated as follows:

$$PSNR = 10.\log_{10}\left(\frac{MAX^2}{MSE}\right) \quad (1)$$

where

$$MSE = \frac{1}{MN}\sum_{x=0}^{M-1}\sum_{y=0}^{N-1}(C(x,y) - S(x,y))^2$$

Note that MAX is the maximum possible pixel value of the images. For example, if the pixels are represented using 8 bits per sample, then the MAX value is 255.

If the stego image has a higher PSNR value, then the stego image has more quality image. Table 1 shows the PSNR value for two stego images in Figures 6 and 7. The PSNR is calculated using the equation of PSNR in Eq. (1).

Based on values of PSNR from Table 1, the PSNR

values show that the stego images have quality images without compromising of the original image.

The pixels of the cover image must fulfill the minimum requirement for the process of data hiding. The minimum image pixel for width is at least 150 while the minimum image pixel for height is at least 112.

Smaller images file size, for example, a BMP image with a sized of 1.0 MB, is proved to be capable of hiding the Secret Message within it. The biggest size of a zipped file to be encoded into a 1.0 MB BMP image by proposed system is 3.16 KB, which means that the size of image can encodes 10553 characters with spaces (or 1508 words or equally to 4 pages of words) underneath the image with near-zero distortion. Both cover and stego images are alike with the images that showed in Fig. 7 with near-zero distortion noticeable by naked eyes. Therefore, the proposed steganography algorithm is a strong yet robust algorithm to produce a stego image which will not be doubted by outsider that the image contains any secret message.

The image file format used in proposed algorithm is focused on bitmap (BMP) format. The BMP file format handles graphics files within the Microsoft Windows OS. Typically, BMP files are uncompressed, hence they are large. The advantage of using BMP files is the simplicity and wide acceptance of BMP files in Windows programs. Thus, this type of image is chosen to be used in our proposed algorithm. Since BMP image has a relatively larger size, the pixels in image are relatively larger as well. Thus, it provides more space for binary codes to be encoded within it. To increase as much as characters that can be hidden, zip technique is used to reduce to total size of file and to enhance the security of the file.

Using the proposed algorithm, we test several sizes of BMP images to see the various sizes of data being stored in the image. Table 2 shows these various results for the testing.

Table 2 shows the comparison of different sizes in BMP image by using the proposed steganography algorithm. These BMP images are used as cover images to encode the zipped file within it. An image is normally contains 3.14 MB. Using the proposed algorithm, the biggest size of a zipped file that can be hidden into and retrieved from a 3.14 MB BMP image is 6.93 KB, which means that the size of image can encodes 27287 characters with spaces (or 4478 words or equally to 10 pages of words) underneath the image with near-zero distortion.

**Table 1** The PSNR value of stego images.

| Image | Reference | PSNR for 1.0 KB embedded inside the image |
|---|---|---|
| Injured Baby | Figure 7 (a) : Stego Image (1) | 76.15 |
| Dog | Figure 7 (b) : Stego Image (2) | 81.47 |

**Table 2** Comparison of different sizes in bitmap images.

| FILE SIZE | | | | Hide Message | Retrieve Message |
|---|---|---|---|---|---|
| Cover Image | Text File | Zipped File | Stego Image | | |
| 438 KB | 4.01 KB | 513 Bytes | 584 KB | √ | √ |
| 438 KB | 12.1 KB | 4.34 KB | Failed | — | — |
| 1.0 MB | 10.4 KB | 3.16 KB | 1.34 MB | √ | √ |
| 1.0 MB | 10.5 KB | 3.15 KB | Failed | — | — |
| 3.14 MB | 12.1 KB | 4.34 KB | 4.19 MB | √ | √ |
| 3.14 MB | 27.0 KB | 6.95 KB | 4.19 MB | √ | √ |
| 3.14 MB | 54.1 KB | 7.03 KB | Failed | — | — |
| 6.74 MB | 54.1 KB | 7.03 KB | 8.99 MB | √ | √ |
| 9.9 MB | 334 KB | 8.48 KB | 13.2 MB | √ | √ |
| 9.9 MB | 335 KB | 8.49 KB | Failed | — | — |

## 5. Conclusions

This paper proposed a new steganography algorithm with 2 layers of security. A system named SIS (Steganography Imaging System) has been developed using the proposed algorithm. We tested few images with various sizes of data to be hidden. With the proposed algorithm, we found that the stego image does not have a noticeable distortion on it (as seen by the naked eyes). We also tested our stego images using PSNR value. Based on the PSNR value of each images, the stego image has a higher PSNR value. Hence this new steganography algorithm is very efficient to hide the data inside the image.

SIS can be used by various users who want to hide the data inside the image without revealing the data to other parties. SIS maintains privacy, confidentiality and accuracy of the data.

## Acknowledgments

This research is supported under the Fundamental Research Grant Scheme (FRGS) Vot 0738.